\documentclass[aps,pra,twocolumn,superscriptaddress]{revtex4-2}

\usepackage[utf8]{inputenc}
\usepackage{soul}
\usepackage{textcomp}
\usepackage[english]{babel}
\usepackage{graphicx}
\usepackage{dcolumn}
\usepackage{bm}
\usepackage{bm}
\usepackage{amsmath}
\usepackage{verbatim}     
\usepackage[dvipsnames]{xcolor}       
\usepackage[normalem]{ulem}   
\usepackage{bbold}
\usepackage{mathrsfs}
\usepackage[mathscr]{eucal}
\usepackage{physics}
\usepackage{braket}
\usepackage{amsmath}
\usepackage{mathrsfs}
\DeclareUnicodeCharacter{0308}{\"}
\newcommand{\beq}{\begin{equation}}
\newcommand{\eeq}{\end{equation}}
\newcommand{\bei}{\begin{itemize}}			
\newcommand{\eei}{\end{itemize}}			

\usepackage{hyperref}
\hypersetup{
   pdfpagemode=None,
   pdfstartpage=1,
   pdfmenubar=true,
   pdftoolbar=true,
   colorlinks = true,
   linkcolor=blue,
   citecolor=blue,
   urlcolor=blue,
   bookmarksopen=false
 }

\newcommand{\fra}[1]{{\color{red}#1}}
\newcommand{\pao}[1]{{\color{purple}#1}}
\begin{document}
\title{Programmable non-Hermitian photonic quantum walks via dichroic metasurfaces}
\author{Paola Savarese}
\affiliation{Dipartimento di Fisica, Universit\`{a} degli Studi di Napoli Federico II, Complesso Universitario di Monte Sant'Angelo, Via Cintia, 80126 Napoli, Italy}
\author{Sarvesh Bansal}
\affiliation{Dipartimento di Fisica, Universit\`{a} degli Studi di Napoli Federico II, Complesso Universitario di Monte Sant'Angelo, Via Cintia, 80126 Napoli, Italy}
\author{Maria Gorizia Ammendola}
\affiliation{Dipartimento di Fisica, Universit\`{a} degli Studi di Napoli Federico II, Complesso Universitario di Monte Sant'Angelo, Via Cintia, 80126 Napoli, Italy}
\affiliation{Scuola Superiore Meridionale, Via Mezzocannone, 4, 80138 Napoli, Italy}
\author{Lorenzo Amato}
\affiliation{PSI Center for Scientific Computing, Theory and Data, CH-5232 Villigen PSI, Switzerland}
\affiliation{Laboratory for Solid-State Physics, ETH Zürich, CH-8093 Zürich, Switzerland}
\author{Raouf Barboza}
\affiliation{Dipartimento di Scienze e Ingegneria della Materia, dell’Ambiente ed Urbanistica, Università Politecnica delle Marche,  Via Brecce Bianche, 60131 Ancona, Italy}
\author{Bruno Piccirillo}
\affiliation{Dipartimento di Fisica, Universit\`{a} degli Studi di Napoli Federico II, Complesso Universitario di Monte Sant'Angelo, Via Cintia, 80126 Napoli, Italy}
\author{Francesco Di Colandrea}
\email{francesco.dicolandrea@unina.it}
\affiliation{Dipartimento di Fisica, Universit\`{a} degli Studi di Napoli Federico II, Complesso Universitario di Monte Sant'Angelo, Via Cintia, 80126 Napoli, Italy}
\author{Lorenzo Marrucci}
\affiliation{Dipartimento di Fisica, Universit\`{a} degli Studi di Napoli Federico II, Complesso Universitario di Monte Sant'Angelo, Via Cintia, 80126 Napoli, Italy}
\affiliation{CNR-ISASI, Institute of Applied Science and Intelligent Systems, Via Campi Flegrei 34, 80078 Pozzuoli (NA), Italy}
\author{Filippo Cardano}\email{filippo.cardano2@unina.it}
\affiliation{Dipartimento di Fisica, Universit\`{a} degli Studi di Napoli Federico II, Complesso Universitario di Monte Sant'Angelo, Via Cintia, 80126 Napoli, Italy}

\begin{abstract}
The evolution of a closed quantum system is described by a unitary operator generated by a Hermitian Hamiltonian. However, when certain degrees of freedom are coupled to an environment, the relevant dynamics can be captured by non-unitary evolution operators, arising from non-Hermitian Hamiltonians. Here we introduce a photonic platform that implements non-unitary quantum walks, commonly used to emulate open-system dynamics, in the synthetic space of light transverse momentum. These walks are realized by propagating light through a series of dichroic liquid-crystal metasurfaces, that impart polarization-dependent momentum shifts. The non-unitary behavior stems from dichroic dye molecules with polarization-dependent absorption, whose orientation is coupled to that of the liquid crystals. We demonstrate multiple walks up to five time steps, with adjustable levels of dichroism set by the metasurface voltage, which is controlled remotely. This discrete-time process maps onto two-band tight-binding models with reciprocal yet non-Hermitian nearest-neighbor couplings, corresponding to a less-studied class of non-Hermitian systems. Our platform broadens the range of optical simulators for controlled investigations of non-Hermitian quantum dynamics.
\end{abstract}
\maketitle
\section*{Introduction}
The physics of non-Hermitian (NH) systems has sparked a growing interest in recent years. Distinctive features are non-orthogonal eigenstates and complex energy spectra, typically embedding exceptional points where the eigenstates coalesce~\cite{ashida2020non}. The landscape of topological phases of these systems is significantly richer than the Hermitian case, with up to 38 symmetry-protected topological classes~\cite{altland1997nonstandard,kawabata2019symmetry}. The breakdown of the conventional bulk-boundary correspondence leads to unique phenomena, such as the NH skin effect, corresponding to a strong localization of the eigenstates towards the boundaries of the system~\cite{RevModPhys.93.015005,revSKIN}. 

Widely used for building quantum algorithms~\cite{PhysRevA.81.042330,doi:10.1126/science.abg7812,reviewquantumwalkcomputing}, modeling transport phenomena~\cite{MARES2020126302}, and simulating topological phases of matter~\cite{kitag}, quantum walks (QWs) constitute the quantum counterparts of classical random walks~\cite{Venegas_Andraca_2012}. 
These processes describe the discrete-time evolution of a particle on a complex lattice, conditioned by an internal degree of freedom, dubbed as the coin. At each time step, the coin undergoes a unitary rotation, followed by a walker displacement, whose direction depends on the coin state. In the simplest scenario, the coin is a two-level system.

QWs provide a convenient framework to investigate NH phenomena. NH topological invariants~\cite{PhysRevLett.119.130501}, emergence of skyrmion patterns~\cite{Wang2019}, light funnelling~\cite{doi:10.1126/science.aaz8727}, quantum phase transitions~\cite{Weidemann2022}, self acceleration~\cite{Xue2024}, and edge burst effects~\cite{PhysRevLett.132.203801} represent examples of intriguing phenomena unique to NH physics that have been recently observed in QW-like experimental settings.

Photonics provides a versatile testbed to engineer NH QWs~\cite{Nasari:23,Wang:23}. 
In optical architectures, the walker's site coordinates can be encoded into different degrees of freedom of light, such as the arrival time of the photons~\cite{PhysRevLett.104.050502}, the beam path~\cite{PhysRevLett.104.153602}, the orbital angular momentum~\cite{doi:10.1126/sciadv.1500087}, and the transverse momentum~\cite{D_Errico_2020}, with the two-level coin typically encoded into light polarization. 
NH photonic lattices can be realized by artificially introducing on-site gains and losses, for instance, exploiting non-linear media~\cite{Chang2014}, partially polarizing optical elements~\cite{PhysRevLett.119.130501,Wang2019}, or by introducing anisotropic couplings between neighboring modes via acousto-optical modulators and variable beam splitters in fiber loops~\cite{doi:10.1126/science.aaz8727,Weidemann2022}.%

In this paper, we present a platform realizing one-dimensional (1D) photonic NH QWs based on structured light propagating through artificially-structured media. This builds on a scheme introduced in Ref.~\cite{D_Errico_2020} for the implementation of unitary QWs on 1D and two-dimensional (2D) lattices, consisting of a light beam propagating in free space, with its polarization and transverse momenta manipulated via a stack of liquid-crystal (LC) metasurfaces. The lattice where the walk takes place is spanned by optical modes corresponding to circularly polarized states of light carrying a quantized amount of transverse momentum, and coin states are the two circular polarization states with opposite handedness. Metasurfaces with a periodic orientation of the LC optic axes, dubbed as $g$-plates~\cite{D_Errico_2020}, give the momentum kick that implements the conditional displacement of the walker. 

Here we show that Hermiticity can be broken by doping the liquid crystals with a dichroic absorbing dye, which features a different absorption for the ordinary and extraordinary linear polarization components~\cite{technical}. The specific directions of the latter are determined point by point by the local optic axis of LC molecules. As detailed below, this enables us to encode NH terms directly in the displacement operator, implemented by $g$-plates, rather than in the  coin rotations. 

In the following, we first model our QW and describe its implementation through dichroic LC metasurfaces. Finally, we report on NH QWs up to five steps in several non-unitary regimes, which are obtained by adjusting both birefrigence and dichroism of each optical element via the application of an oscillating electric field. 

\section*{Results}
\subsection{Photonic implementation}
\begin{figure}[!tp] 
    \includegraphics[scale=0.43]{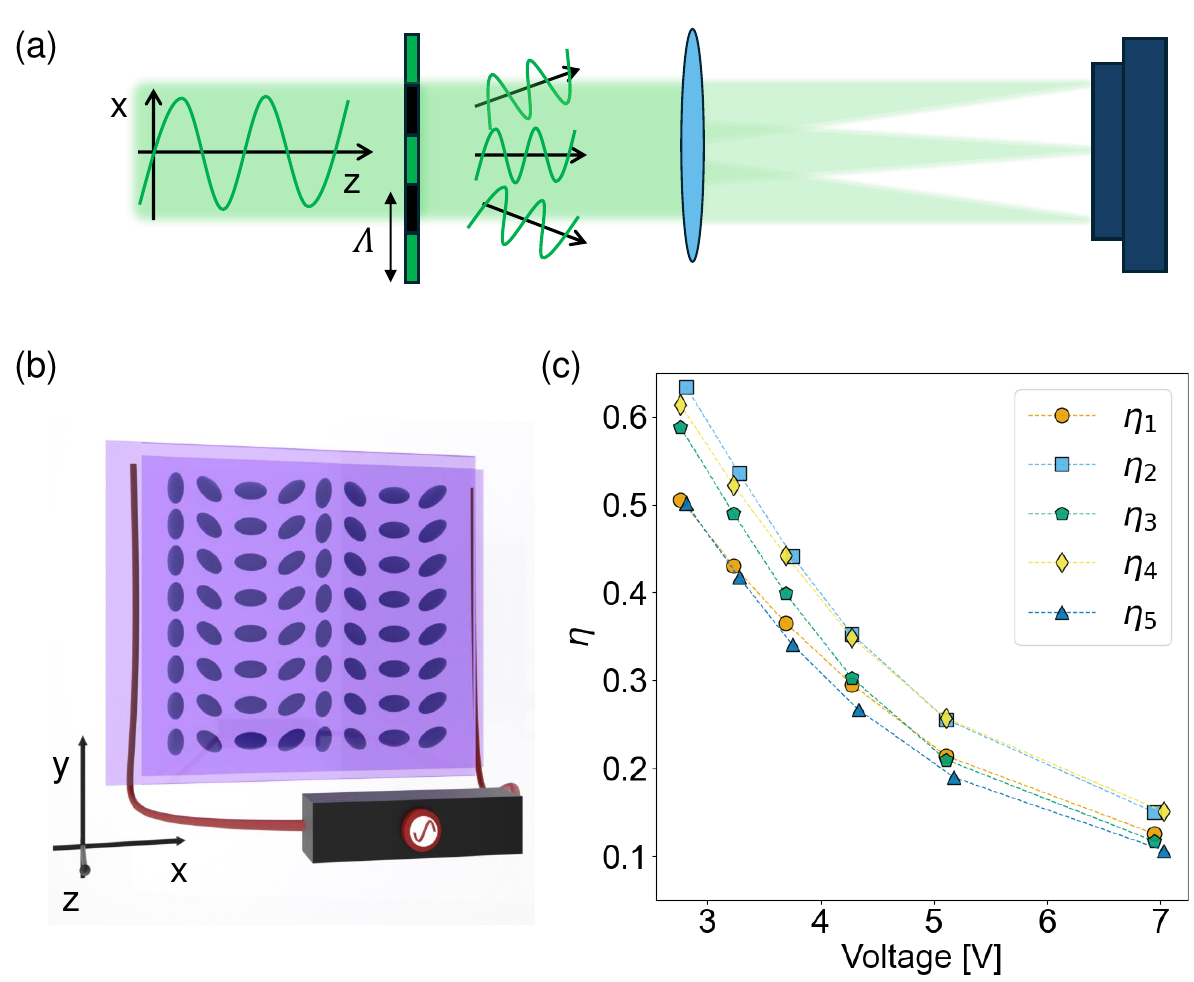}
    \caption{\textbf{Transverse momentum modes and dichroic $g$-plates}. (a)~Optical modes for the QW simulation are copropagating Gaussian beams with a slightly tilted propagation direction. The spatial period of their transverse phase profile is $\Lambda/m$, where $m$ is the mode index. A $g$-plate with spatial period ${\Lambda}$ is used to create and manipulate these modes, which can be sorted in the focal plane of a lens. (b)~In dichroic metasurfaces, the dye molecules align with the LC substrate. Accordingly, the application of an electric field allows for simultaneously tuning the birefringence and the dichroic power of the device. (c)~Measured values of the parameter ${\eta}$ at six voltages corresponding to ${\delta = \pi \pmod{2\pi}}$ for five dichroic ${g\text{-plates}}$.
    }
 \label{fig:concept}
\end{figure}

The expression of the optical modes $\ket{m,\phi}$ spanning the lattice sites of our QWs reads~\cite{D_Errico_2020}
\begin{equation}\label{eq:modes}
    \ket{m,\phi}=A(x,y,z)e^{i k_{z} z} e^{i m\Delta k x}\ket{\phi},
\end{equation}
where $A(x,y,z)$ is a Gaussian envelope, $k_z$ is the longitudinal wavevector component, and $\ket{\phi}$ represents the polarization (coin) state, being left-circular ($\ket{L}$) or right-circular ($\ket{R}$), respectively. These modes propagate in free space and carry a quantized amount of transverse momentum given by $m \Delta k$, where $m$ is an integer number and $\Delta k$ is the quantum of transverse momentum imparted to the modes at each step. ${\Delta k}$ is set to be larger than the transverse momentum spread of the Gaussian envelope with beam waist ${w_0}$, i.e., ${w_0\geq 2\pi/\Delta k}$~\cite{D_Errico_2020}.
By keeping ${m\Delta k\ll k_z}$, modes in Eq.~\eqref{eq:modes} correspond to overlapped Gaussian beams propagating along the $z$ axis with a slightly tilted wavefront in the $x$ direction, which can be spatially resolved on a CCD camera placed in the focal plane of a lens (see Fig.~\ref{fig:concept}(a)).
QW dynamics have been realized by propagating the modes in Eq.~\eqref{eq:modes} through stacks of LC polarization gratings ($g\text{-plates}$)~\cite{D_Errico_2020} having spatial period ${\Lambda=2\pi/\Delta k}$, recently generalized to more sophisticated devices in one~\cite{Di_Colandrea_2023} and two~\cite{10.1117/1.AP.7.1.016006} spatial dimensions. In our implementation, we set ${\Lambda=5\text{ mm}}$. In the circular polarization basis, their action on modes defined in Eq.~\eqref{eq:modes} reads
\begin{equation}
    T_\delta= 
    \begin{pmatrix}
    \cos{\frac{\delta}{2}} & i \sin{\frac{\delta}{2}} \hat t\\[5pt]
    i \sin{\frac{\delta}{2}} \hat t^\dagger & \cos{\frac{\delta}{2}}
    \end{pmatrix},
    \label{eqn:g-plate}
\end{equation}
where ${\hat t\ket{m,j}=\ket{m- 1,j}}$ ($\hat t^\dagger\ket{m,j}=\ket{m+ 1,j}$). The optical birefringence $\delta$, controlling the hopping amplitude between neighboring sites, is uniform across the device but can be electrically tuned~\cite{Piccirillo2010,Rubano:19}. These devices shift the walker by giving a transverse momentum kick, whose sign depends on the input polarization, thus effectively realizing a coin-dependent translation operator on the lattice. The coin rotation can also be implemented with homogeneous LC plates acting as standard quarter-wave plates:
\begin{equation}
W=\frac{1}{\sqrt{2}}\begin{pmatrix}
    1 & i\\
    i & 1
\end{pmatrix}.
\label{eqn:coinrotation}
\end{equation}

\begin{figure*} 
    \includegraphics[scale=0.55]{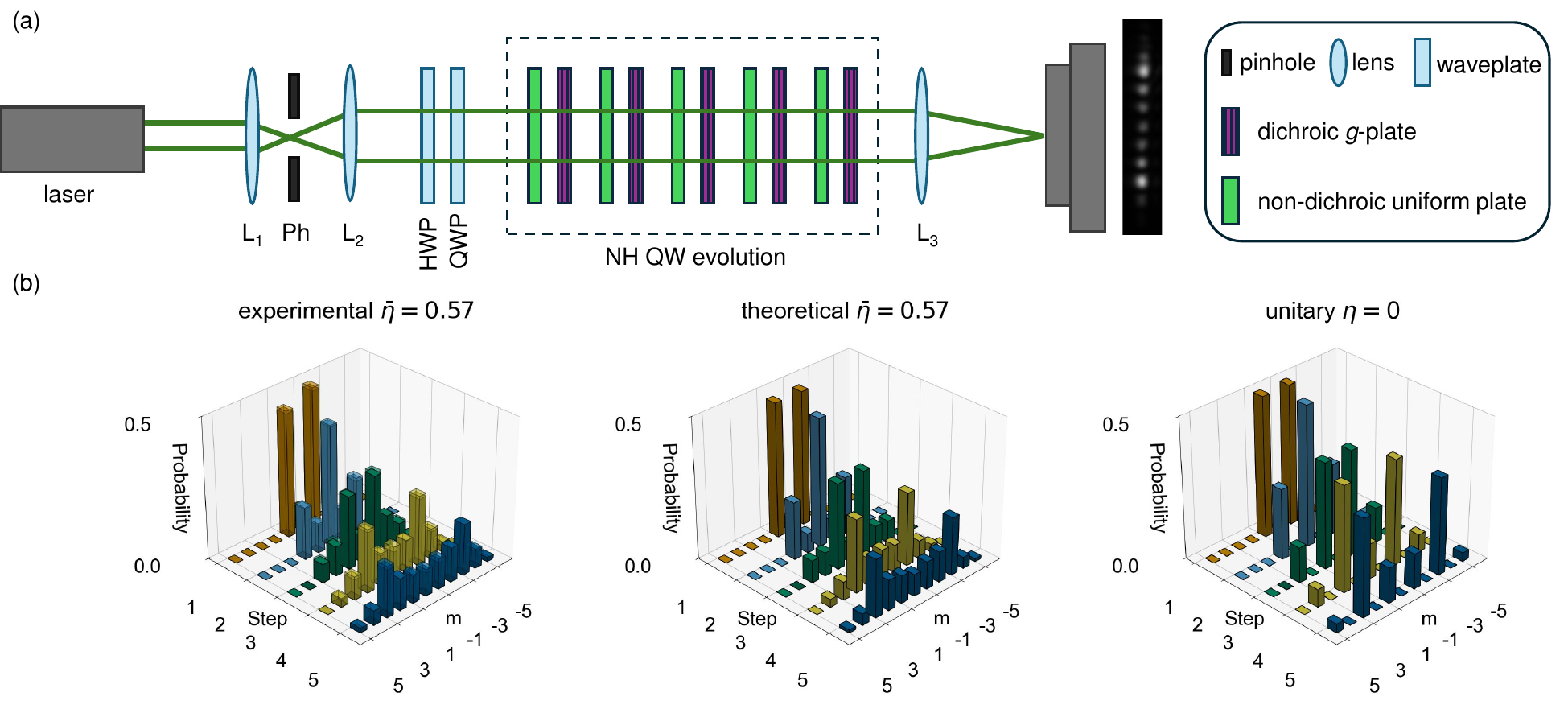}
    \caption{\textbf{Experimental implementation.} (a)~Sketch of the experimental setup for the simulation of NH QWs. A Gaussian beam is expanded and spatially filtered with lenses ($\text{L}_1$-$\text{L}_2$) in a telescopic configuration and a pinhole (Ph). Its polarization is adjusted with a half-wave plate (HWP) and a quarter-wave plate (QWP), which corresponds to setting the coin input state. It then propagates through a sequence of dichroic LC metasurfaces and uniform plates, whose optical actions realize the operations of the NH QW dynamics. The output modes can be resolved in the focal plane of a lens, where they form a line of separated Gaussian spots, from which the experimental probability distribution can be extracted. (b)~Representative experimental and theoretical QW distributions corresponding to a protocol with ${\delta=\pi}$, average dichroism ${\bar\eta=0.57}$, and horizontal input polarization, $\ket{H}=(\ket{L}+\ket{R})/\sqrt 2$. Experimental results (left) are compared to numerical simulations of the implemented QW (center) and of the unitary QW having the same value of $\delta$ but $\eta=0$ (right).}
 \label{fig:setup}
\end{figure*}
In this work, we focus on the QW protocols obtained by alternating operators $T_\delta$ and $W$. Given an input state $\ket{\psi_0}$, the evolved state after $t$ time steps is computed as $\ket{\psi(t)}=U^t\ket{\psi_0}$, where $U$ is the single-step operator $U=T_\delta W$. This setting realizes a unitary evolution.

The key idea of the scheme presented here is to introduce non-unitary elements into the propagation via dichroism, that is, polarization-dependent absorption, in addition to standard birefringence. This is achieved by adding dichroic molecules, specifically 1,4-Diaminoanthraquinone, to the LC mixture of our devices~\cite{Marrucci2000, technical}. This organic dye features linear dichroism, that is, it absorbs light differently depending on the orientation of the optical polarization. Using the Dirac notation, light transmission through a thin film of this material can be modeled as ${e^{-\eta_o}\ketbra{o}+e^{-\eta_e}\ketbra{e}}$, where $\ket{o}$ ($\ket{e}$) is the local ordinary (extraordinary) eigenpolarization, and $\eta_o$ ($\eta_e$) is the corresponding absorbance, typically with ${\eta_e>\eta_o}$~\cite{Marrucci2000,guesthost_top}. When mixed with LC, dye molecules align along the local optic axis~\cite{guesthost_top}. In the case of devices with patterned orientation of the latter, also the dichroic axes inherit the same spatial structure (see Fig.~\ref{fig:concept}(b)).

By combining the dye response with the $g$-plate operator in Eq.~\eqref{eqn:g-plate}, we obtain a generalized non-unitary displacement operator~\cite{technical}
\begin{equation}
    T_{\delta,\eta}= e^{-\frac{\eta'}{2}}
    \begin{pmatrix}
    \cos{\frac{\delta+i\eta}{2}} & i \sin{\frac{\delta+i\eta}{2}} \hat t\\[5pt]
    i \sin{\frac{\delta+i\eta}{2}} \hat t^\dagger & \cos{\frac{\delta+i\eta}{2}}
    \end{pmatrix},
    \label{eqn:generalized}
\end{equation}
where ${\eta=\eta_{e}-\eta_{o}}$ is the dichroic parameter, while ${\eta'=\eta_{e}+\eta_{o}}$ is a global attenuation factor, that is irrelevant for the system dynamics and can therefore be neglected.

Due to the coupling between dye and LC molecules, the application of an external electric field allows us to simultaneously access the dichroic parameter and the birefringence~\cite{technical}, thus accessing an entire family of non-unitary evolutions. Qualitatively, we expect both $\eta$ and $\delta$ to decrease when increasing the field strength, eventually vanishing for strong fields. However, due to the periodic nature of the dependence of the displacement operator on $\delta$ (see Eq.~\eqref{eqn:generalized}), multiple voltages lead to the same effective birefringence yet to a different dichroism. In this paper, we focus on the case ${\delta = \pi \pmod{2\pi}}$, corresponding to half-wave retardation, which is obtained for six different voltage values for the five metasurfaces used in our experiment, as detailed below. The corresponding measured values of ${\eta}$ are reported in Fig.~\ref{fig:concept}(c), with average values ${\bar\eta=(0.57,0.48,0.40,0.31,0.23,0.13)}$. The procedure for extracting $\eta$ from polarimetric measurements~\cite{technical} is detailed in the Methods. Slight deviations within the batch of plates can be ascribed to small variations in the concentration of the dye molecules in the LC mixture and the overall device impedance. In particular, we notice that the measured dichroism for $g$-plates 1 and 5 is systematically lower than the others. 
\newline\indent Finally, we note that our walk models a 1D NH system with left and right nearest-neighbor hopping parameters that are equal in magnitude yet are not complex conjugates of each other (see Eq.~\eqref{eqn:generalized}).
This configuration differs from previous experimental studies reporting either non-reciprocal couplings~\cite{doi:10.1126/science.aaz8727,Weidemann2022}, yielding the NH skin effect, or alternated gain and losses on consecutive lattice sites~\cite{PhysRevLett.119.130501,Chang2014}, thus providing an interesting addition to photonic simulators of NH dynamics. 

\begin{figure*}[!tp]                
\centering

    \includegraphics[width=1\linewidth]{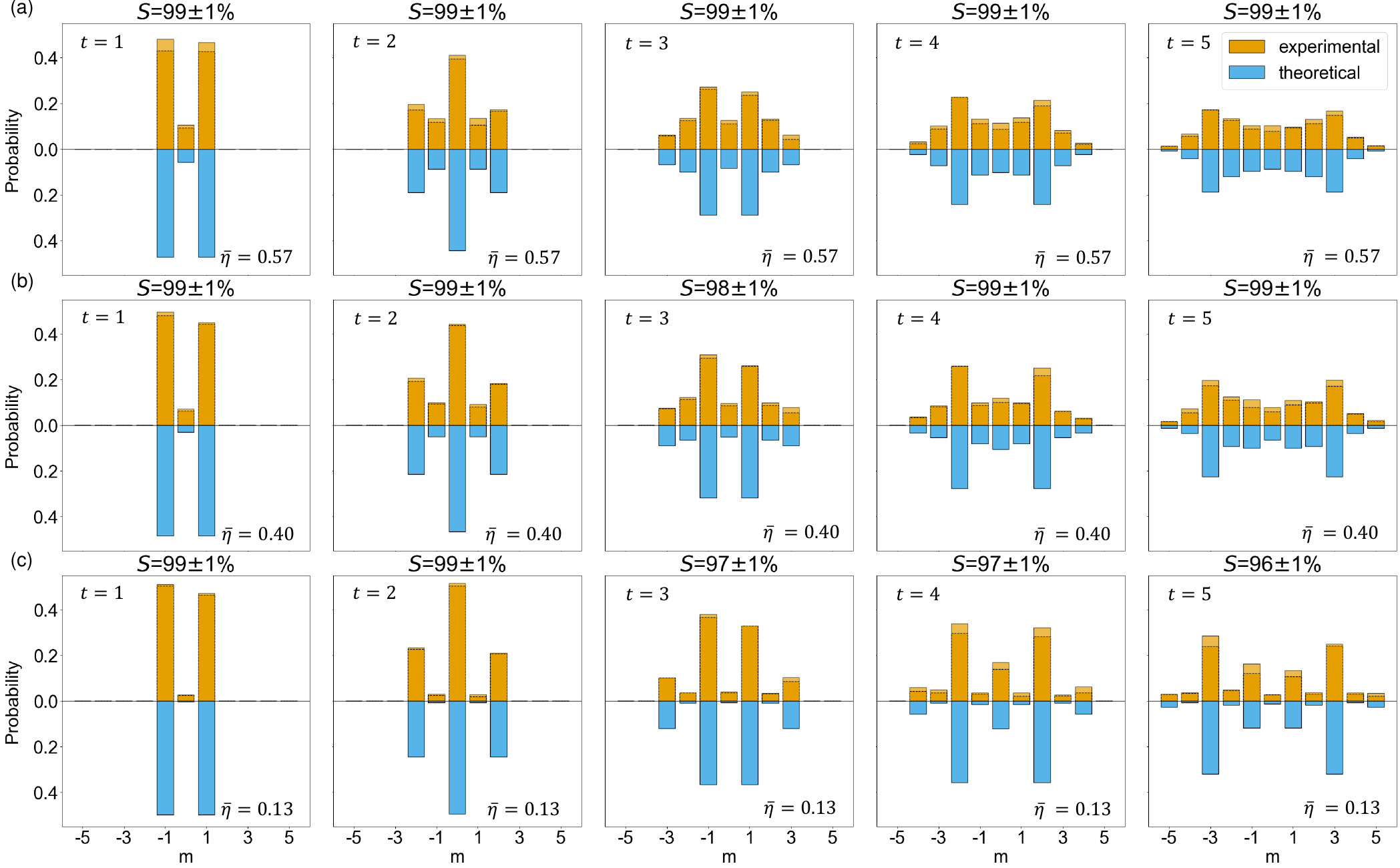} 
     \caption{\textbf{Experimental results.} Evolution of a localized walker in a NH QW up to five steps. Experimental data and numerical simulations refer to horizontal polarization as the input coin state, with different levels of average dichroism: (a)~${\bar\eta=0.57}$, (b)~$\bar\eta=0.40$, (c)~$\bar\eta=0.13$. The average similarity obtained at each step is reported within each panel. The dashed horizontal lines represent the mean distribution for the experimental data and the shadowed region above each bar corresponds to one standard deviation.}
    \label{expdata2}
\end{figure*}
\subsection{Experimental results}
Non-unitary QWs are realized through the experimental setup sketched in Fig.~\ref{fig:setup}(a). A 532-nm linearly polarized laser is magnified through a telescope system (made of two lenses with focal
lengths ${\text{L}_{1}=35\text{ mm}}$ and ${\text{L}_{2}=50\text{ mm}}$). A pinhole, placed in the focal plane, operates as a spatial filter. The beam waist at the output of the second lens is measured to be ${w_0\simeq 5 \text{ mm}}$. A half-wave plate (HWP) and a quarter-wave plate (QWP) set the input coin-polarization state. The beam then propagates through dichroic $g$-plates and uniform unitary plates, implementing the coin-dependent (non-unitary) translation and the coin rotation, respectively, up to 5 time steps. These devices are located in a compact box provided with external screws, which allows adjusting their transverse displacement. At the exit of the box, the beam passes through a lens (${\text{L}_{3}=500 \text{ mm}}$) implementing an all-optical Fourier transform, which sorts the output modes in the focal plane over a straight line, each corresponding to a different walker site. Light intensity is recorded with a CCD camera and the power of each Gaussian spot ${I_m}$ is computed by integrating the relative intensity. After normalization with respect to total light power, we obtain the occupation probability of each lattice site: ${P_m=I_m/\sum_{m'} I_{m'}}$. In Fig.~\ref{fig:setup}(b), we plot the experimental distribution obtained for a walker initialized on a single site when all the $g$-plates are tuned at the first voltage associated with half-wave retardation, corresponding to an average dichroism $\bar\eta=0.57$. Numerical simulations are computed by setting for each $g$-plate in the sequence the corresponding value of $\eta$ reported in Fig.~\ref{fig:concept}(c). A comparison with the unitary walk ($\eta=0$) at ${\delta=\pi}$ helps reveal the effect of non-unitary dynamics on the output state distribution, which tends to be less spread towards the edges compared to the unitary case.
In Fig.~\ref{expdata2}, we plot the experimental and theoretical distributions obtained for three voltage values corresponding to $\bar\eta = 0.57, 0.40,$ and 0.13, respectively, and for a horizontally polarized input state.
The agreement with the theoretical predictions is quantified in terms of the similarity estimator:
\begin{equation}
    S=\left(\sum_{m}\sqrt{P_{\text{exp}}(m)P_{\text{th}}(m})\right)^{2},
\end{equation}
where $P_{\text{exp}}(m)$ and $P_{\text{th}}(m)$ are the experimental and 
theoretical probability distributions, respectively. All experimental data are in good agreement with the theory, as witnessed by the large values of the reported similarity. Each data point is computed as averages over four independent measurements, with error bars given by the standard deviations. Experimental results for other values of dichroism are provided in the Supplementary Material.



\begin{figure}[h!]
\centering
    \includegraphics[scale=0.4]
    {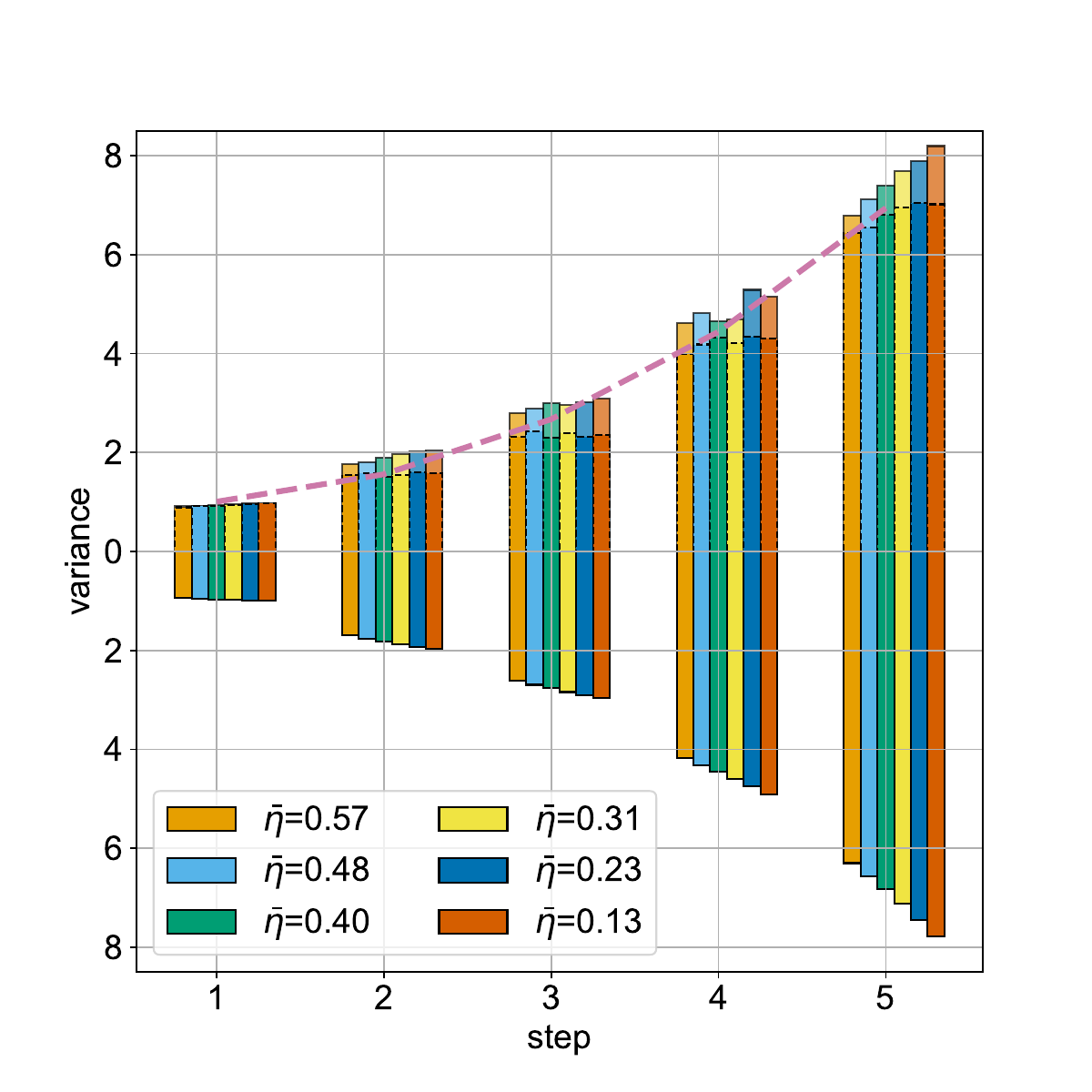}
    \caption{\textbf{Walker variance in a NH QW.} Variance of the walker probability distribution as a function of the number of steps. The input polarization is $\ket{H}$. Experimental data (top) are compared with theoretical prediction (bottom). The shadowed region above each bar corresponds to one standard deviation. The pink dashed curve is obtained as the average best fit of the variance for the different configurations at each step and serves as a guide for the eye.}
    \label{fig:variance}
\end{figure}

These results certify that weak dissipative regimes are associated with output distributions exhibiting near-zero probabilities for intermediate walker sites, which mostly resembles the unitary case, while for higher values of the dichroic parameter, the distribution tends to spread over all available sites. Nevertheless, the overall walker distribution preserves the typical ballistic spreading of unitary walks~\cite{Venegas_Andraca_2012}, as shown in Fig.~\ref{fig:variance}, where we plot the experimentally retrieved variance as a function of the step number for a horizontally polarized input state and the six different values of dichroism reported above, also compared with numerical simulations. This is expected since the presence of dichroism does not act as a source of decoherence in the evolution, which would lead to stronger localization phenomena, as already demonstrated for walks featuring spatial~\cite{Crespi2013} and temporal disorder~\cite{PhysRevLett.123.140501}.
\newline
\section*{Conclusion}
We developed a bulk-optical platform to simulate the non-unitary evolution of a single particle on a dissipative lattice system. The photonic simulation relies on optical losses that are induced by an organic dichroic compound added to the traditional LC devices. By adjusting the external voltage, the platform can be programmed to explore different dissipation regimes, simulating different levels of interaction with an external environment.

In this work, we focused on the simulation of specific QW dynamics in the presence of losses. In the near future, our setup could be engineered to probe NH topological phases~\cite{Nasari:23,yan2023advances} and measure the quantum geometry of NH models~\cite{e15093361,PhysRevB.103.125302,PhysRevResearch.6.023202}, also exploring multi-photon regimes~\cite{Wangclassquant,PhysRevResearch.6.023216}.

\section*{Methods}
\subsection*{Extraction of the dichroic parameter $\eta$}
When a dichroic compound with high-order parameter is incorporated into a host LC matrix, its molecules align with the LC director. This is known as the \emph{guest-host} effect~\cite{guesthost_top}. This implies that the application of an external field induces a torque on both LC and dichroic molecules, allowing for simultaneous control of both birefringence and dichroism. In the case of dichroic metasurfaces with uniform optic-axis orientation, the dependence of the dichroic parameter on the external voltage can be characterized by recording the transmitted power for the ordinary and extraordinary waves at different voltages:
\begin{equation}
\begin{split}
I_\text{ord}&=e^{-2\alpha_od}I_\text{0},\\
I_\text{ext}&=e^{-2\alpha_ed}I_\text{0},
\end{split}
\end{equation}
where $I_0$ is the total light power, $\alpha_o$ and $\alpha_e$ are the ordinary and extraordinary absorption coefficients, respectively, and $d$ is the sample thickness (approximately equal to $17$ $\mu$m in our plates). The dichroic parameter ${\eta=(\alpha_e-\alpha_o)d}$ can be determined from
\begin{equation}
\eta=\frac{1}{2}\log{\frac{I_\text{ord}}{I_\text{ext}}}.
\end{equation}
In the case of spatially patterned devices, such as dichroic $g$-plates, this result holds only for a sufficiently small region across the sample, where the optic-axis modulation can be regarded as uniform. For this reason, a lens with a focal length of $10$~cm is employed to focus the beam on a small area of the device, approximately covering a portion of $\Lambda/500$.

In our experiment, a square-wave signal at approximately 4 KHz was used. As shown in Fig.~\ref{fig:concept}(c), increasing the voltages induces a larger out-of-plane tilt of the LC molecules, which eventually become parallel to the propagation direction, resulting in turn in a significant reduction of the dichroic response of the cell. Further details can be found in Ref.~\cite{technical}.
\bibliography{bibliography}
\bigskip
\noindent\textbf{Acknowledgments.}
This work was supported by the PNRR MUR project PE0000023-NQSTI.
\bigskip
\newline \noindent \textbf{Disclosures.}
The authors declare no conflicts of interest.
\bigskip
\newline \noindent \textbf{Data availability statement.}
The data that support the findings of this study are available from the corresponding authors upon reasonable request.

\clearpage
\onecolumngrid
\renewcommand{\figurename}{\textbf{Figure}}
\setcounter{figure}{0} \renewcommand{\thefigure}{\textbf{S{\arabic{figure}}}}
\setcounter{table}{0} \renewcommand{\thetable}{S\arabic{table}}
\setcounter{section}{0} \renewcommand{\thesection}{S\arabic{section}}
\setcounter{equation}{0} \renewcommand{\theequation}{S\arabic{equation}}
\onecolumngrid

\begin{center}
{\Large Supplementary Material for: \\Programmable non-Hermitian photonic quantum walks via dichroic metasurfaces}
\end{center}
\vspace{1 EM}

\section*{Supplementary Data}
We provide experimental results for a five-step NH QW with different values of the average dichroic parameter: $\bar\eta=0.48,0.31$, and 0.23. Horizontal polarization is used as input state.
\begin{figure*}[!h]                
\centering

    \includegraphics[width=1\linewidth]{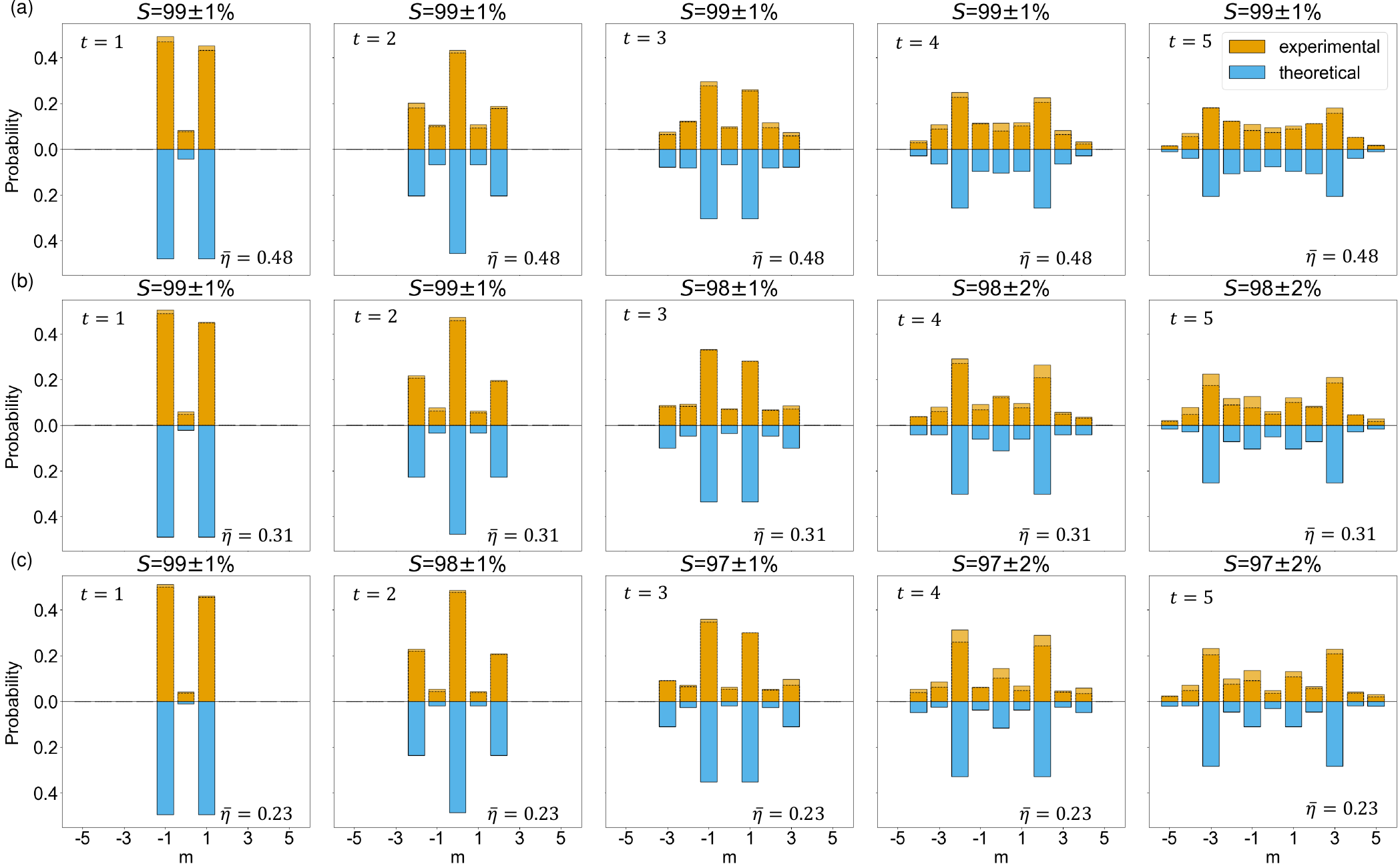} 
     \caption{\textbf{Experimental results.} Evolution of a localized walker in a NH QW up to five steps. Experimental data and numerical simulations refer to horizontal polarization as the input coin state, with different levels of average dichroism: (a)~${\bar\eta=0.48}$, (b)~$\bar\eta=0.31$, (c)~$\bar\eta=0.23$. The average similarity obtained at each step is reported within each panel. The dashed horizontal lines represent the mean distribution for the experimental data and the shadowed region above each bar corresponds to one standard deviation.}
    \label{expdata3}
\end{figure*}

\end{document}